\documentclass[letterpaper,nofonttune]{IEEEtran}
\usepackage{xpatch}
\makeatletter
\patchcmd\@makecaption{\\}{.\,~}{}{\fail}
\patchcmd{\@makecaption}
  {\scshape}
  {}
  {}
  {}
\def\tablename{Table}
\makeatletter
\IEEEoverridecommandlockouts
\usepackage{color}
\usepackage{cite}
\usepackage{amsmath,amssymb,amsfonts}
\usepackage{algorithmic}
\usepackage{graphicx}
\usepackage{textcomp}
\usepackage[switch]{lineno}

\usepackage[ruled,linesnumbered,noend]{algorithm2e}
\SetAlFnt{\small}
\usepackage{multirow}
\usepackage[dvipsnames]{xcolor}
\usepackage{mathtools}
\usepackage{bm}
\usepackage{microtype}
\usepackage{pmboxdraw}
\usepackage{listings}
\usepackage{siunitx}
\usepackage{mathptmx}
\usepackage{booktabs}
\usepackage{enumitem}
\usepackage[caption=false]{subfig}

\def\BibTeX{{\rm B\kern-.05em{\sc i\kern-.025em b}\kern-.08em
    T\kern-.1667em\lower.7ex\hbox{E}\kern-.125emX}}

\DeclarePairedDelimiterX\set[1]{\lbrace}{\rbrace}{ #1}

\newcommand\restr[2]{{\left.\kern-\nulldelimiterspace #1 \right\rvert_{#2}}}
\newcommand\restrr[2]{{\kern-\nulldelimiterspace #1 \rvert_{#2}}}

\newcommand{\lp}{\left(}
\newcommand{\rp}{\right)}

\newcommand{\lV}{\lVert}
\newcommand{\rV}{\rVert}

\newcommand\grad{\nabla}

\newcommand{\ncandidate}{\ensuremath{n_\text{candidate}}}
\newcommand{\ninteraction}{\ensuremath{n_\text{interaction}}}
\newcommand{\threshold}{\ensuremath{r_\text{cut}}}
\newcommand{\twall}{\ensuremath{t_\text{wall}}}

\newcommand{\metatext}[1]{}

\newcommand{\Ninterior}{\ensuremath{N_\mathrm{interior}}}
\newcommand{\Nghost}{\ensuremath{N_\mathrm{ghost}}}
\newcommand{\Nnode}{\ensuremath{N_\mathrm{atom}}}

\title{Breaking the Molecular~Dynamics Timescale Barrier Using a Wafer-Scale System
\thanks{Corresponding authors emails: michael@cerebras.net, srajama@sandia.gov}
}

\author{%
\IEEEauthorblockN{%
Kylee Santos\IEEEauthorrefmark{1},
Stan Moore\IEEEauthorrefmark{2},
Tomas Oppelstrup\IEEEauthorrefmark{3},
Amirali Sharifian\IEEEauthorrefmark{1},
Ilya Sharapov\IEEEauthorrefmark{1},
Aidan Thompson\IEEEauthorrefmark{2}, \\
Delyan Z Kalchev\IEEEauthorrefmark{1},
Danny Perez\IEEEauthorrefmark{4},
Robert Schreiber\IEEEauthorrefmark{1},
Scott Pakin\IEEEauthorrefmark{4},
Edgar A. Leon\IEEEauthorrefmark{3},
James H Laros III\IEEEauthorrefmark{2}, \\
Michael James\IEEEauthorrefmark{1},
and
Sivasankaran Rajamanickam\IEEEauthorrefmark{2}}\\
\IEEEauthorblockA{\IEEEauthorrefmark{1}Cerebras Systems, Sunnyvale, CA} \\
\IEEEauthorblockA{\IEEEauthorrefmark{2}Sandia National Laboratories, Albuquerque, NM} \\
\IEEEauthorblockA{\IEEEauthorrefmark{3}Lawrence Livermore National Laboratory, Livermore, CA}\\
\IEEEauthorblockA{\IEEEauthorrefmark{4}Los Alamos National Laboratory, Los Alamos, NM}}

\begin{document}
%\linenumbers

\maketitle

\begin{abstract}

Molecular dynamics (MD) simulations have transformed our understanding of the nanoscale, driving breakthroughs in materials science, computational chemistry, and several other fields, including biophysics and drug design.  Even on exascale supercomputers, however, runtimes are excessive for systems and timescales of scientific interest.
Here, we demonstrate strong scaling of MD simulations on the Cerebras Wafer-Scale Engine.
By dedicating a processor core for each simulated atom, we demonstrate a 179-fold improvement in timesteps per second versus the Frontier GPU-based Exascale platform, along with a large improvement in timesteps per unit energy.
Reducing every year of runtime to two days unlocks currently inaccessible timescales of slow microstructure transformation processes that are critical for understanding material behavior and function.

Our dataflow algorithm runs Embedded Atom Method (EAM) simulations at rates over 270,000 timesteps per second for problems with up to 800k atoms. This demonstrated performance is unprecedented for general-purpose processing cores.

\end{abstract}


\section{Overview of the Problem}
\label{section:introduction}

\begin{figure}[tp]
\centerline{\includegraphics[bb=0 0 3.75in 3in]{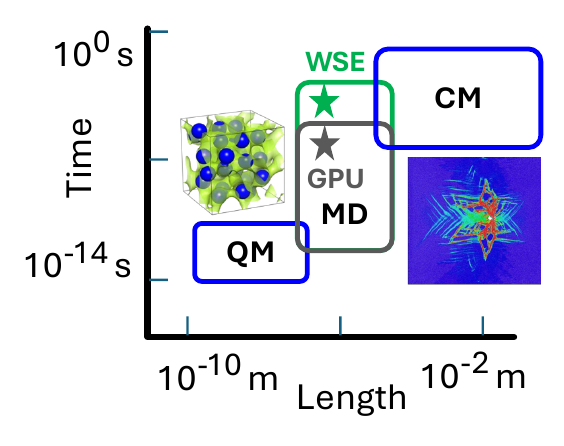}}
\caption{Comparison of maximum MD timescale achievable using Cerebras Wafer-Scale Engine (WSE, green) and Exascale GPU hardware (GPU, gray). The boxes represent typical achievable ranges of length and time using different materials simulation approaches: quantum electronic  methods (QM, left box), molecular dynamics (MD, middle box) and continuum mechanics (CM, right box).  Green and gray stars reflect measured performance for 800,000 Ta atoms (see Fig.~\ref{fig:perfpanel}), assuming 30 days of wall-clock time on WSE and GPU hardware, respectively. The nearly 180-fold increase in maximum achievable timescale for MD using WSE is transformative for a broad range of applications in materials science, chemistry, and physics.
}
\label{fig:MDScales}
\vspace*{-14pt}
\end{figure}

In today's massively parallel distributed memory systems, the natural scaling path for physics simulations is weak scaling.
Weak scaling enables scientists to simulate increasingly large systems, but it does nothing to address \emph{timescale limitations}. Timescale limitations are caused by a breakdown in strong scaling due to latencies within the HPC system.

In conventional supercomputers, the gaps between compute speed and inter-node communication bandwidth and latency are huge and growing.  To accommodate this, simulations frequently pack ever-more work onto each node and then scale out simulation size with node count. This is weaker than weak scaling, and it makes for increasing runtime to cover an interesting simulation timescale.

Classical molecular dynamics (MD) provides an important bridge between the length and time scales of quantum electronic methods and those of continuum mechanics methods (Fig.~\ref{fig:MDScales}).
MD methods are particularly impacted by timescale limitations. On one hand, they must resolve atomic vibrations with femtosecond timestepping; on the other hand, a broad range of important physical phenomena in materials science, physics, and chemistry emerge only at much longer timescales, e.g., on the order of 100~microseconds for annealing of radiation damage in nuclear reactors, thermally activated catalytic reactions, phase nucleation close to equilibrium, protein folding, etc.

Whereas weak scaling has enabled enormous billion-atom~\cite{carbon_GB} and twenty-trillion-atom~\cite{tchipev2019twetris} simulations as well as high-accuracy simulations via methods with 10,000x higher computational intensity than traditional models~\cite{deepmd_GB,allegro_GB,Plimpton2012},
even month-long exascale runs can, at most, encompass just a few microseconds of simulated time.
Typical simulation rates saturate around 1,000 timesteps per wall-clock second or less.
This limits the sequential steps in a month-long MD simulation to no more than the number of milliseconds in a month, about 3 billion. Only improved strong scaling can enable simulation of the necessary timescales. Thus, we seek at least a hundredfold, and ultimately a thousandfold, improvement in timesteps per second.

To break the MD timescale barrier, one needs a highly parallel system with communication bandwidth comparable to compute throughput, and communication latency comparable to core clock frequency. This could allow extreme strong scaling to only one atom per processor. The advent of wafer-scale computing, in particular the Cerebras Wafer-Scale Engine (WSE), which embeds nearly one million processors and a communication and memory system with the desired performance on a single silicon wafer, promises to support efficient strong scaling and a large gain in timesteps per second for systems of interesting size. 

Grain boundaries are regions where atomic crystal lattices of different orientation meet. They are ubiquitous in any macroscopic piece of metal and have profound effects on material characteristics, including strength, heat tolerance, and corrosion resistance. Phase transformation and other atomic processes in grain boundaries are very slow at most conditions, and scientific understanding is severely limited by accessible simulation time scales. In Tokamak fusion reactors technological progress is limited by our understanding of grain-boundary evolution. There, the vessel containing and facing the fusion plasma typically is made of tungsten due to its high melting point. Unfortunately, machinability and durability of tungsten are severely limited by its brittleness, which relates to grain-boundary behavior~\cite{frolov2018}. For this reason, we choose reference simulations relevant to elucidation of grain-boundary behavior.

\begin{figure}[tp]
\centering
\subfloat{\includegraphics[width=0.45\textwidth, bb=0 0 18.8in 5.5in]{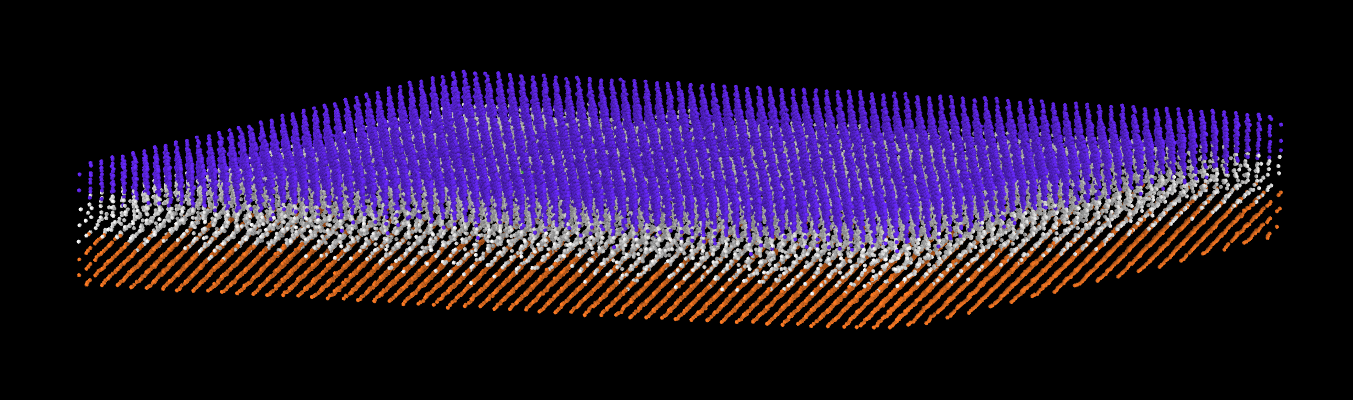}}\\
\subfloat{\includegraphics[width=0.45\textwidth, bb=0 0 18.8in 5.5in]{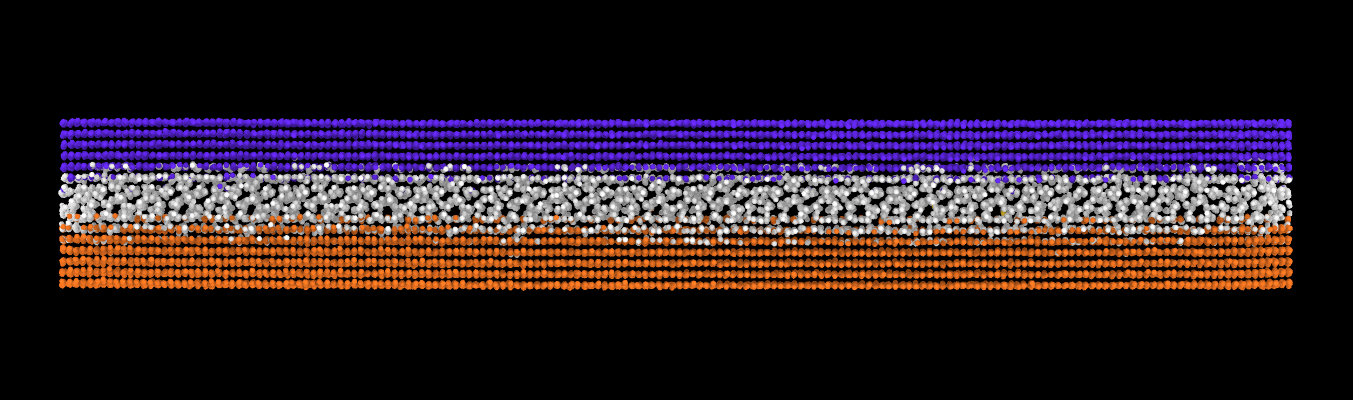}}
\caption{Two views of a grain boundary in tungsten (W). Atoms in the grain boundary are shown in white. The other colors represent different crystal orientations. Upper panel: the difference in crystal lattice orientation can be seen above and below the grain boundary. Lower panel: Although more complex and less clearly defined, there is also structure in the grain boundary.} 
\label{fig:GB}
\vspace*{-10pt}
\end{figure}

Thin-slab structures efficiently model grain boundaries. Atoms far from the interface region have perfect crystalline organization, while atoms proximal to the grain boundary form complex and slowly evolving structures (Fig.~\ref{fig:GB}). Open (non-periodic) boundary conditions allow atoms to migrate in and out on the edges of the slab as driven by phase change processes in the grain boundary \cite{GrainBoundary,frolov2018}. 

This paper reports on an effort to achieve orders-of-magnitude acceleration of MD for physically relevant systems, thereby dramatically expanding the timescale horizon amenable to direct MD simulation. Key contributions include:
\begin{enumerate}[wide]
    \item A method for efficiently strong scaling MD simulations on the Cerebras Wafer-Scale Engine, thereby greatly extending accessible MD timescales by over 100x,
    \item A simple performance model which is nonetheless accurate to within 3\% of the achieved performance, and
    \item Demonstrated simulations on 800,000 atom metallic crystal lattices similar to that shown in Fig.~\ref{fig:GB} (but without defects) running at 274,000 timesteps per second, corresponding to a 179-fold speedup over Frontier.
\end{enumerate}
                   % III
\section{Current State of the Art} % IV
\label{section:overview}

\subsection{Computational Molecular Dynamics}

MD simulations model atoms as points that obey Newton's laws of motion. Trajectory integration divides force by mass to get each atom's acceleration $\vec{a}(t)$. The laws of motion dictate that the atom's trajectory $\vec{r}(t)$ satisfy the ordinary differential equation
\begin{equation}\label{eq:motion}
\vec{r}\,'' = \vec{a}.
\end{equation}

A conservation law requires a force acting on the system to be the negative gradient of an \emph{interatomic potential} energy that depends on the atoms' positions. The choice of potential is a modeling decision. For long timescale insight we seek a potential that reproduces material behaviors while demanding minimal computational intensity. This rules out strictly pairwise potentials such as Lennard-Jones~\cite{LJ} because they are unable to replicate basic crystal properties. The criteria also exclude the more expensive potentials that incorporate quantum physics training data through machine learning. Empirical many-body potentials therefore offer a favorable speed/accuracy tradeoff for long time-scale simulations. 

The Embedded Atom Method (EAM)~\cite{EAM1,EAM2,EAM3,EAM4} is an empirical many-body potential comprising pairwise interatomic components and many-body electronic embedding components.
In many cases it reproduces structures found in real materials and/or through accurate quantum-mechanical calculations.

EAM maintains a set of $N$ atoms that contribute to an electron density field $\rho(\vec{r})$. Each atom $i$ exists at a point $\vec{r}_i$ and makes an isotropic contribution $\rho_i$ to density so that
\begin{equation}\label{eq:density}
\rho(\vec{r}) = \sum_i \rho_i ( \lV \vec{r} - \vec{r}_i \rV ).
\end{equation}
Each atom has an embedding energy $F_i$
that depends non-linearly on the electron density at its position. Every pair of atoms $(i, j)$ has an interaction potential $\phi_{ij}$ that depends on their Euclidean distance $r_{ij} \equiv \lV \vec{r}_j - \vec{r}_i \rV$. The system’s potential energy is given by
\begin{equation}\label{eq:potential}
U = \sum_{i\ne{j}} \frac12\phi_{ij}(r_{ij}) + \sum_i F_i(\rho(\vec{r}_i)).
\end{equation}
Note that the density, force, and potential functions $\rho_i$, $F_i$, and $\phi_{ij}$ are atom-dependent, allowing for heterogeneous ensembles of atoms. 
Our implementation allows them to be arbitrary polynomial splines based on atom type.

Density $\rho_i$ and interatomic potentials $\phi_{ij}$ decay rapidly. In practice, they are made to vanish exactly beyond a threshold interaction-cutoff radius \threshold. In this way, the sums in~\eqref{eq:density} and~\eqref{eq:potential} are evaluated only for pairs of atoms located within a distance \threshold~from each other.

The potential's negative gradient, $-\grad U = -\partial U / \partial \vec{r}_i $, provides the force acting on the system. 
Isotropic potentials and densities cause force contributions to act radially in the direction $\bar{r}_{ji} \equiv (\vec{r}_i - \vec{r}_j){r_{ji}^{-1}}$:
\begin{equation}\label{eq:EAMforce}
\begin{aligned}
\grad U \equiv \frac{\partial U}{\partial \vec{r}_i} = \sum_{\substack{j\ne i \\ r_{ij} < \threshold}} 
\Bigl[ & F_i'(\rho(\vec{r}_i))\rho_j'(r_{ij}) +{} \\
       & F_j'(\rho(\vec{r}_j))\rho_i'(r_{ji}) + \phi_{ij}'(r_{ij})  \Bigl] \bar{r}_{ji}.
\end{aligned}
\end{equation}

We use the Verlet leap-frog scheme to convert acceleration into discrete-time velocity $\vec{v}^{\;(t)}$ and position $\vec{r}^{\;(t)}$ along a dynamical trajectory. Leap-frog integration approximates position at each timestep and velocity at the mid-points between timesteps:
\begin{equation}\label{eq:leapfrog}
\begin{split}
\vec{v}^{\;\lp k+\frac{1}{2}\rp} &= \vec{v}^{\;\lp k-\frac{1}{2} \rp} + \vec{a}^{\;(k)}\, \Delta t,\\
\vec{r}^{\;\lp k+1\rp} &= \vec{r}^{\;\lp k \rp} + \vec{v}^{\;\lp k+\frac{1}{2} \rp}\, \Delta t.
\end{split}
\end{equation}
This results in a second-order explicit method for the equations of motion; the method preserves net momentum and angular momentum and is time-reversible and symplectic.
Symplecticity means that the integration recovers the dynamics of some Hamiltonian that approximates the exact Hamiltonian of the atom system. This makes the simulation physically meaningful even over very long simulation times.         % IV-A
\label{section:sota}

\subsection{Quantitative Performance}

Achieving direct long-timescale simulations is one of the most pressing challenges facing the materials-modeling community, and for systems as large as those we consider (on the order of a million atoms) it is largely out of reach today because of the difficulty of strong scaling.

Even for the computationally lightweight Lennard-Jones (LJ) potential~\cite{LJ} and a very small system size of only 1k atoms (mimicking the strong scaling limit) on an NVIDIA V100 GPU, the max timestepping rate of the LAMMPS production code was reported at less than 10k timesteps/s~\cite{LAMMPS}. This is due to kernel-launch overhead, although recent kernel-fusion work sped up this case by $\sim$20\%~\cite{kernel_fusion_PR}. Strong-scaling to multiple GPUs adds even more kernel launches as well as MPI communication overhead. The situation on a dual-socket Intel Skylake CPU (using 36 MPI ranks) is slightly better, achieving $\sim$25k timesteps/s for the 1k atom LJ system \cite{LAMMPS}. It is very likely that MPI communication cost dominates for this small system size, limiting the timestepping rate on the CPU.

When the goal is simulation of $O(10^{-4})$ seconds of real time at a timestep on the order of a few femtoseconds, the needed $O(10^{11})$ timesteps can be accomplished in the roughly $10^5$ seconds in a day only with million-timestep-per-second performance. Thus for this situation, neither CPU- or GPU-based systems can get close to the required simulation rate, regardless of the node count of the machine.

There are alternatives to strong scaling MD simulations; methods such as parallel replica dynamics~\cite{PRD} can provide long-timescale simulations. However, these methods apply only to specific classes of systems and certain conditions.

Specialized rather than general-purpose hardware is another possibility.
The bio-science community has pursued such machines, for instance MDGRAPE \cite{morimoto2021hardware} and Anton \cite{anton_GB}. \mbox{Anton-3} is a 512-node fixed-point computer using various custom-width integer representations in its hardened datapaths. \mbox{Anton-3's} maximum reported timestepping rate is $9.8 \times 10^{5}$ timesteps/s for 24k atoms and $7.7 \times 10^{5}$ timesteps/s for 328k atoms. To the best of our knowledge, \mbox{Anton-3} is intended for modeling systems of biological interest and has not been used for material science.

We note that MD codes commonly compute forces using FP32 (single precision) to improve performance but may use FP64 (double precision) for other operations such as trajectory integration because this minimizes accumulated error over many timesteps~\cite{gromacs,gromacs2,LAMMPS}.
                       % IV-B
\section{Innovations Realized}
\label{section:mapping}

\subsection{Locality-preserving Atom Mapping}\label{sec:localmapping}

\begin{figure*}
\centering
\subfloat[][]{\includegraphics[width=0.15\textwidth, bb=0 0 3.31in 3.31in]{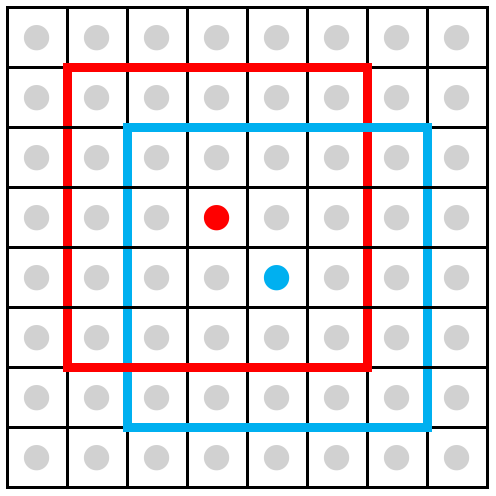}\label{fig:neighborhood}}\;
\subfloat[][]{\includegraphics[width=0.15\textwidth, bb=0 0 2.61in 2.62in]{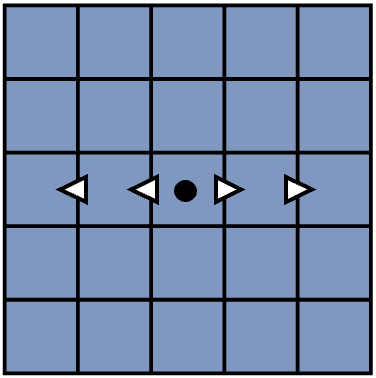}\label{fig:horizontal}}\;
\subfloat[][]{\includegraphics[width=0.15\textwidth, bb=0 0 2.61in 2.64in]{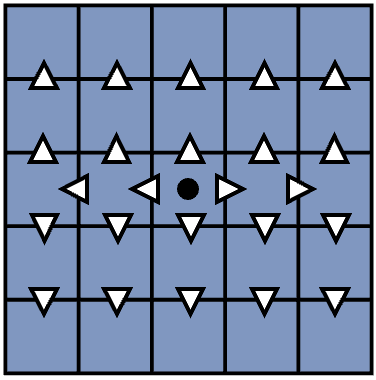}\label{fig:horizontalvertical}}\;
\subfloat[][]{\includegraphics[width=0.15\textwidth, bb=0 0 1.64in 1.64in]{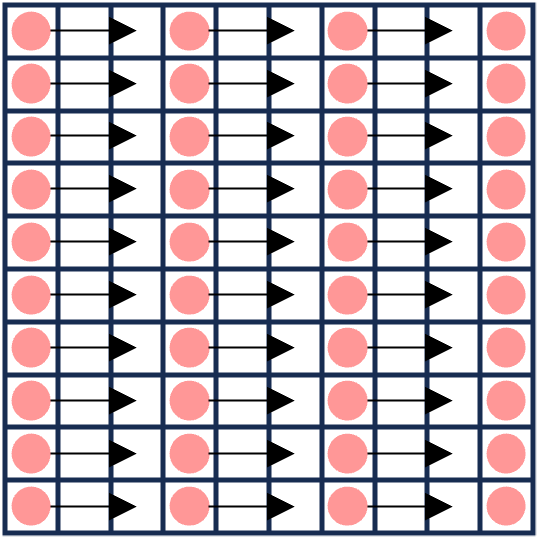}\label{fig:horizontalsystolic1}}\;
\subfloat[][]{\includegraphics[width=0.15\textwidth, bb=0 0 1.64in 1.64in]{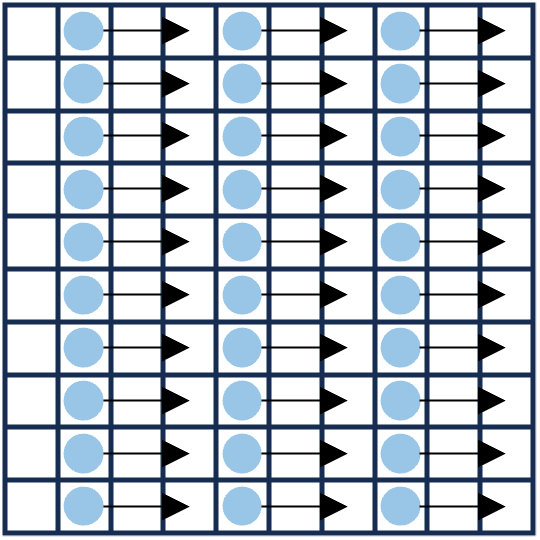}\label{fig:horizontalsystolic2}}\;
\subfloat[][]{\includegraphics[width=0.15\textwidth, bb=0 0 7.5in 7.5in]{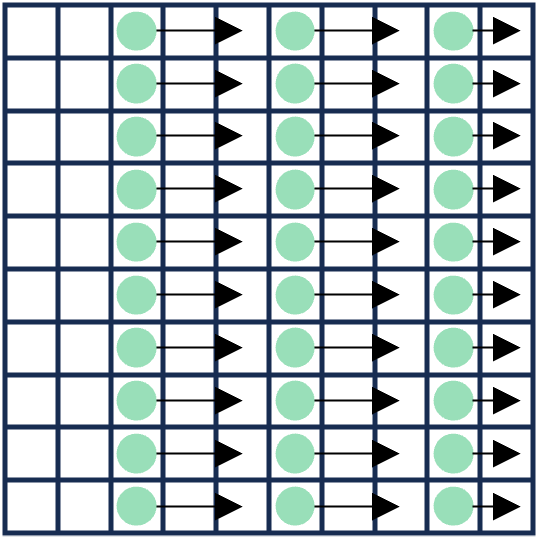}\label{fig:horizontalsystolic3}}
\caption{Example of the candidate exchange for a $5\times 5$ neighborhood, i.e. $b=2$.
\textbf{(a)}~Overlapping neighborhoods of two distinct atoms. All atoms within the red and blue atoms' interaction thresholds are contained within the red and blue outlined regions, respectively.
\textbf{(b)}~Horizontal stage of neighborhood multicast
\textbf{(c)}~Vertical stage of neighborhood multicast
\textbf{(d)}~First positive-horizontal-multicast transmission. Red head tiles multicast atom data to right two tiles
\textbf{(e)}~Second multicast transmission. The roles of tiles in the multicast domain have shifted one tile to the right.
\textbf{(f)}~Last multicast transmission. After this, all tiles in the fabric have transmitted their atom two hops to the right. Leftward transmission (not depicted) occurs concurrently. 
}
\vspace*{-7pt}
\end{figure*}

In this work, we develop an EAM-based MD algorithm for mesh-connected architectures that integrate independent cores on a 2D Cartesian grid. We then implement the algorithm on the Cerebras Wafer-Scale Engine (WSE) platform (Sec.~\ref{section:wse}).

In our test case examples, the number of cores is slightly larger than the number of atoms.  We therefore map atoms to cores so that the mapping is one-to-one.  Let $g(i)$ be the core to which atom $i$ is mapped.  Thus,
each core within the processor acts as a \emph{worker} responsible for a single atom. It maintains the atom’s identity $i$, position $\vec{r_i}$, and velocity $\vec{v_i}$. It also stores local copies of interpolation tables for $\rho_i$, $F_i$, and $\phi_{ij}$.

At each timestep, workers must exchange the positions and velocities of interacting atoms.  We choose mappings of atoms to cores such that interacting atoms are mapped to close mesh neighbors.
To do so, we identify the array of cores with the base of the simulation domain so that each core $c$ has a nominal $(x,y)$ coordinate, which we denote $P(c)$. In this way, each core is associated with the column of space above it. A projection $P$ flattens the simulation domain onto its x-y plane (zeroing the atoms' z-coordinate). The \emph{assignment cost} $C(g)$ of an assignment $g$ of atoms to cores is the worst-case coordinate displacement between $P(\vec{r_i})$ and $P(g(i))$. The cutoff threshold \threshold~and assignment cost determine the fabric distance 
separating the worker cores of interacting atoms, which is bound by $2C(g) + \threshold$.

We accomplish data exchange for interacting atoms by exchanging atom data between all cores whose max norm distance is less than a core-independent value $b$. Thus, at runtime we set $b$ so that every $(2b+1)$-wide square neighborhood of fabric contains all interactions for the atom at the neighborhood's center.

From the viewpoint of a generic core $c = g(i)$, which is the worker integrating the motion of atom $i$, a timestep proceeds as follows:
\begin{enumerate}[wide]
\item \textbf{Candidate exchange.} Core $c$  multicasts its atom's identity and position to its square of neighboring cores (in the box of extent $2b+1$) and receives the corresponding information from these neighbors. (Sec.~\ref{sec:candidateexchange}).
\item \textbf{Neighbor list.} With all potentially close atoms $j$ in memory, $c$ computes $r_{ij}^2$ (no need to take the square root) and discards atoms that lie beyond the cutoff using $r^2_\text{cut}$ (Sec.~\ref{ssec:neighborlist}). 
\item \textbf{Embedding calculation and exchange.} $c$ uses the neighbor atoms within $\threshold$ to compute its embedding energy $F_i$. This energy is also communicated with neighbors because it contributes to force observed by other atoms. (Sec.~\ref{sec:candidateexchange}).
\item \textbf{Force calculation and integration.} Now $c$ has all terms for evaluating ${\partial U}/{\partial \vec{r_i}}$. It calculates the acceleration at timestep $k$, updates its atom's velocity (at timestep $k + 1/2$), then updates its position $r_i$ (at timestep $k+1$) according to the Verlet leap-frog scheme. 
\item \textbf{Atom swap.} To maintain interaction locality in the presence of moving atoms, cores may swap atoms with a neighbor (Sec.~\ref{sec:atomswap}).
\end{enumerate}

\vspace{-10pt}
\subsection{Neighborhood Exchange}\label{sec:candidateexchange}
Interacting atoms share tiny amounts of information with each other at two parts of the timestep. 
At the start of the timestep, atoms exchange positions consisting of three scalar coordinates (12 bytes); later in the timestep they exchange scalar embedding energies (4 bytes). 

Every core has a
unique neighborhood (Fig.~\ref{fig:neighborhood}) that partially overlaps to varying extents with other neighborhoods.  The locality-preserving mapping of atoms to cores ensures that the neighborhoods of the worker cores of interacting atoms always overlap.

Therefore, in the algorithm's first step, workers in every neighborhood participate in an all-to-all exchange of their atom data. Because the traffic from different cores can share mesh links, the messages must interleave with each other as they are disseminated through their overlapping multicast domains. This subsection details the novel communication pattern that orchestrates this.

Workers communicate concurrently using multicast messages to avoid redundant fabric traffic. This communication pattern uses consecutive horizontal (Fig.~\ref{fig:horizontal}) and vertical (Fig.~\ref{fig:horizontalvertical}) stages. In the horizontal stage, information about one atom propagates a distance $b$ to the right and left; in the vertical stage, the accumulated data from $2b+1$ atoms propagates up and down a distance $b$. 

When $b > 1$, the data from multiple sources must use the same mesh links.  To prevent link contention, we orchestrate the transmit and receive roles of the stage into $b+1$ short phases, in each of which a subset of the cores sends, so as not to compete for the use of the mesh links.
To do this, our implementation configures the routers and cores to use a novel systolic \emph{marching multicast} communication scheme.
It partitions the worker grid into non-overlapping vertical strips of size $(b+1)\times n_y$, where $n_x$ and $n_y$ denote the dimensions of the wafer's mesh. In each of $b+1$ phases, one of the cores in each row of every strip multicasts, and all of these do so in parallel~(Fig.~\ref{fig:horizontalsystolic1}-\subref*{fig:horizontalsystolic3}). The first transfer multicasts from the first column of each strip across its width. Each subsequent transfer does the same with the multicast domain shifted one tile to the right, until all columns in the strip are visited. The transfers are staged in this way so that they do not conflict for communication-mesh links.

Two virtual channels are used in the horizontal stage; two others are used in the vertical stage.  A distinct virtual channel is used for positive- and a second for negative-axis directed traffic.
Each core runs four parallel threads.  On each virtual channel there is one send and one receive thread.  The threads progress in parallel, with hardware tracking the state, sending data whenever the outgoing link becomes available, and receiving and storing incoming data when it arrives.

Each thread's send or receive is programmed with a single vector move instruction.  When sending, a memory vector is sent to a ``fabric" vector.   When receiving, the direction is from a fabric vector to a memory buffer.
Each transmit thread follows the move with an additional instruction to transmit a command \emph{wavelet} (a single-word fabric message) on the channel, indicating to the fabric routers
the completion of the current vector transmissions, which on arrival at a router triggers a change of the router state for the next phase of the stage.

\newsavebox{\routerlisting}
\begin{lrbox}{\routerlisting}%
\begin{minipage}{0.32\textwidth}
\vspace{-4.5cm}
\begin{small}
\begin{lstlisting}[
    keywords={parallel,serial},
    literate={%
      {∀}{\ensuremath{\forall}}{1}
      {←}{\ensuremath{\gets{} }}{1}
      {in}{\ensuremath{\in{} }}{1}
    },
    columns=fullflexible,
    frame=tlrb
  ]
parallel {
    serial {
      lr[] ← atom;
      lr[] ← {(ADV, ADV, RST), (ADV)};
    }
    serial {
      rl[] ← atom;
      rl[] ← {(ADV, ADV, RST), (ADV)};
    }
    ∀j in [0,b+1) row[j]   ← lr[];
    ∀j in [0,b+1) row[j+b] ← rl[];
}
\end{lstlisting}
\end{small}
\end{minipage}
\end{lrbox}

\begin{figure*}
\centering
\newlength{\iww}
\settowidth{\iww}{\includegraphics[bb=0 0 9.5in 5.5in]{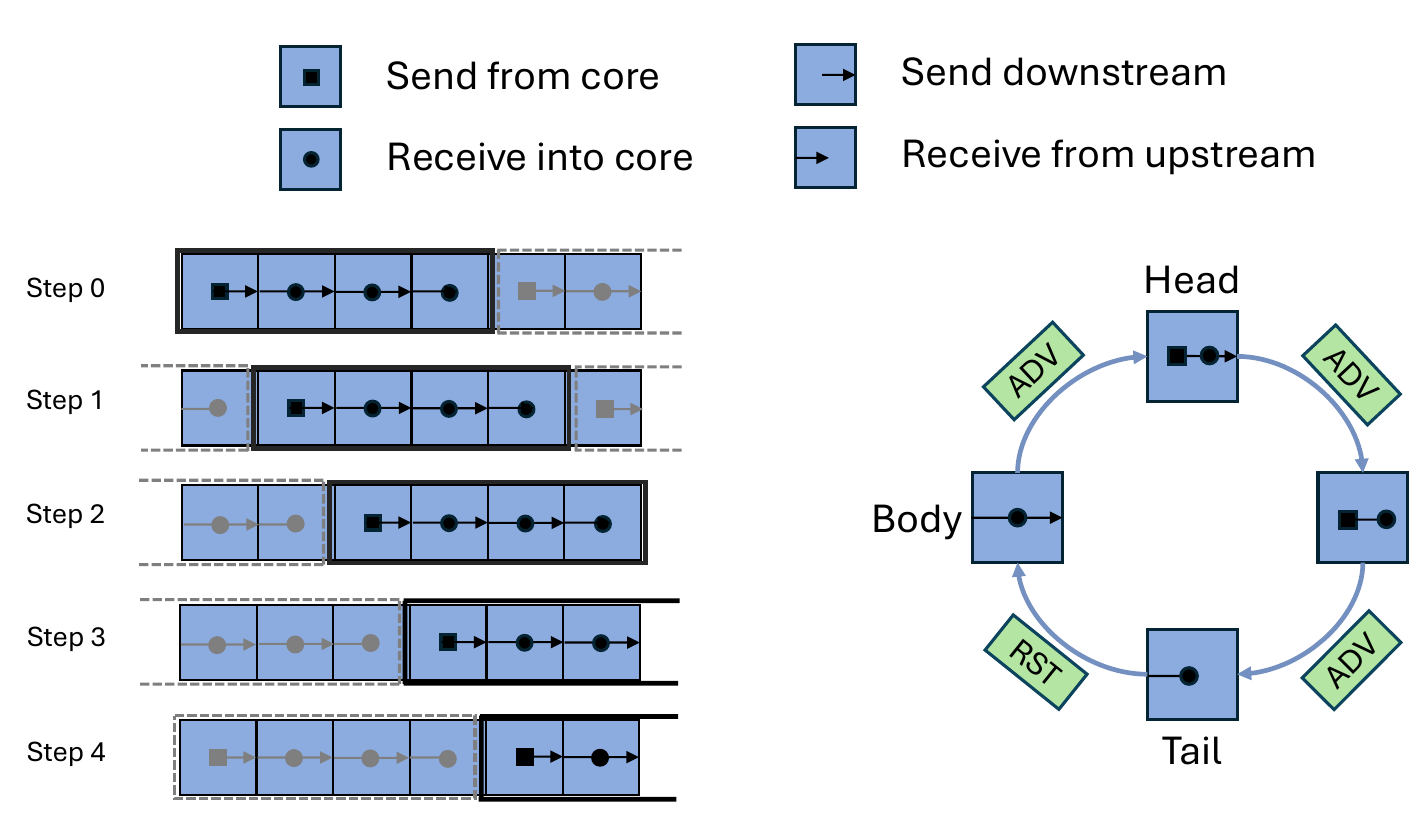}}
\subfloat[][]{\includegraphics[trim={0 0 0.5\iww{} 0},clip,width=0.27\textwidth, bb=0 0 4.75in 5.5in]{figures/exchange.pdf}\label{fig:marching-pipeline}}
\subfloat[][]{\includegraphics[trim={0.5\iww{} 0 0 0},clip,width=0.27\textwidth, bb=4.75in 0 9.5in 5.5in]{figures/exchange.pdf}\label{fig:marching-states}}\quad\quad
\subfloat[][]{\usebox{\routerlisting}\label{lst:horiz-stage}}
\caption{\textbf{(a)}~Systolic routing pipeline diagram for the marching multicast, with $b=3$. \textbf{(b)}~Router state machine for marching multicast. \textbf{(c)}~Tungsten code for neighborhood communication's horizontal stage. The vertical stage differs only in its transfer size.}
\label{fig:marching-multicast}
\vspace*{-7pt}
\end{figure*}

The router alternates between three states during the systolic pipeline (Fig.~\ref{fig:marching-pipeline}).
A router accepts data from its local core only when it is in the head state.
The next $b-1$ tiles are in the body state. They receive data from upstream and multicast it
to both their local core and downstream. The tail state is the last tile in the head's
multicast neighborhood; it forwards incoming messages to its local core.

Each router uses pre-configured rules for head, body, and tail states. When the head completes
its transmission, it sends a command to the routing plane that causes the head to proceed to the
tail state; the next tile in line to proceed to the head state; and the tail to proceed to the
body state. Hardware constraints disallow simultaneously changing a router's input and output,
so the actual router state machine uses four states (Fig.~\ref{fig:marching-states}). Control wavelets contain lists of router commands. Based on their configuration, routers can react to the first command in the list and/or pop the first command from the list before forwarding to downstream routers. The marching multicast command wavelets instruct tiles to ``advance'' to the next state or to ``reset'' to body state. The body tiles are
configured to pop ``advance'' commands so that only the first body tile in the chain reacts to the
head's ``advance'' command.

Fig.~\ref{lst:horiz-stage} presents the code for the horizontal stage.  It is written in a WSE-domain-specific language that, coincidentally, is named \emph{Tungsten}.

\vspace{-8pt}
\subsection{Neighbor list}\label{ssec:neighborlist}

The atoms received during candidate exchange are a superset of the atoms that workers interact with.
In the neighbor-list step, workers compute the squared distance between their atom and every candidate atom.
They build a \emph{neighbor list} of atoms that are within the cutoff range.
The candidate atoms arrive in a deterministic order, so the neighbor list is trivially a list of ordinal numbers of admitted candidates.
The workers immediately use the neighbor list to gather candidates into contiguous memory.
This facilitates vectorized operations in subsequent steps.

\vspace{-8pt}
\subsection{Atom Swap}\label{sec:atomswap}

An occasional greedy remapping counteracts the effect of atom motion.
The procedure uses two neighborhood exchanges. First, cores exchange atom state and calculate the change in assignment cost for all swaps they could participate in. Then, cores exchange the identifier of their best swap partner. When a core detects a mutual agreement of swap preference, it overwrites its local atom state. Empty tiles are allowed (represented as atoms at infinity) to facilitate freedom of reassignment.

\vspace{-8pt}
\subsection{Periodic Boundary Conditions (PBC)\label{sec:pbc}}
\begin{figure}[tp]
\centerline{\includegraphics[width=0.75\linewidth]{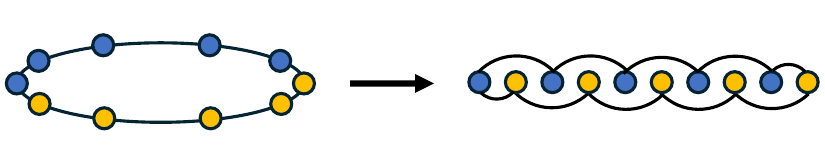}}
\vspace*{-7pt}
\caption{Periodic space to a line segment.}
\label{fig:periodic}
\vspace*{-7pt}
\end{figure}

The mapping naturally maintains locality for $z$-direction periodicity. Periodicity in $x$ or $y$ requires care to avoid long-distance communication. The approach is to split the coordinate circle of a periodic dimension in two parts and collapse it to a line (Fig.~\ref{fig:periodic}). Atoms from the two sides of the circle interleave on the wafer, keeping interacting atoms near each other.

Consequently, communicating workers are two hops away instead of one hop in the non-periodic case.
While expanded neighborhoods increase the on-chip data transfers, bandwidth is not our limiting resource (Sec.~\ref{sec:FabricLoadPBC}). 
                % V
\section{How Performance Was Measured}
\label{section:performance_criteria}

\subsection{Wafer-Scale Engine}
\label{section:wse}
We measure performance of our algorithm on the Cerebras Wafer-Scale Engine (WSE) platform.
WSEs are monolithic processors constructed as the largest square possible within a (circular) 300~mm silicon wafer. These 46,225~mm\textsuperscript{2} chips are the world's largest and, in terms of peak performance, also the most powerful. We describe aspects of the processor relevant to this work and refer readers to~\cite{WSE_HPC_2,HotChips} for additional details.

\begin{figure}[tp]
\centerline{\includegraphics[width=0.4\textwidth]{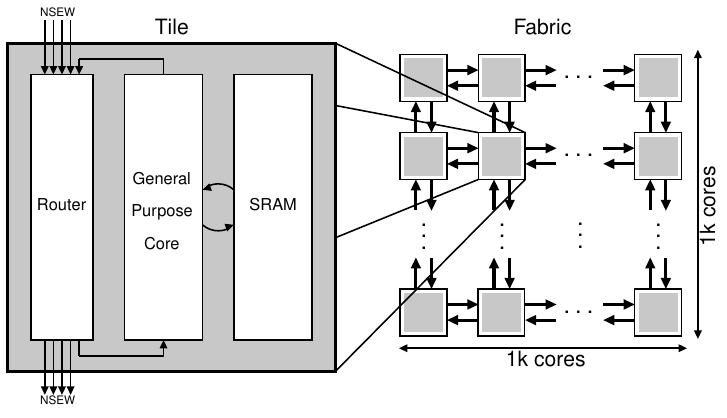}}
\caption{Wafer-scale engine as a fabric of tiles. Each tile has its own general purpose core, 48 kB of SRAM, and a router connecting it to its four neighbors.}
\label{fig:tile}
\vspace*{-7pt}
\end{figure}

The WSE comprises a Cartesian grid of \emph{tiles}~(Fig.~\ref{fig:tile}), each with a general-purpose processor core, dataflow fabric router, and primary memory. The WSE-2 processor has 850,000 cores in a ${\sim}920{\times}920$ array and draws 23~kW of power. Tiles are mesh-connected to their four nearest neighbor tiles on the wafer. A single configurable oscillator synchronously clocks the entire array. Hardware mechanisms transparently correct fabrication defects so that the design maintains uniform topology, uniform bandwidth, and guaranteed latency throughout the processor.

The WSE-2 provides a total of 40~GB of single-cycle latency, on-chip SRAM\@.  This is distributed across tiles (48~kB apiece) and the local memory is the only tile-accessible memory.  There is no cache hierarchy.  Local memory is physically addressed.  The cumulative memory bandwidth is 20~PB/s. At the tile level, memory bandwidth matches datapath bandwidth, so that binary operations do not encounter a memory wall: they can sustain two operand reads for every result written. Each 64-bit wide datapath can perform four 16-bit operations or two 32-bit operations every cycle from non-contiguous memory locations. Tiles communicate using single-word messages (or vectors of these) over the dataflow fabric.  Bandwidth is one message per cycle independent of the vector length.  Message send and receive are expressed in the instruction set of the tile's processor, with routing, queuing, and flow control done in hardware and with no message passing software layer.  The upshot is extremely high communication bandwidth and extremely low communication latency.

The fabric routers allow the exchange of up to ten 32-bit messages on every cycle: one in each direction to the local core and to all four neighboring routers. All physical channels support 24 virtual channels, each having a dedicated routing table and link level buffers. In aggregate, the fabric delivers 20~PB/s of interconnection bandwidth. The latency between neighboring routers is one cycle. Generally, confining application communication to localized regions of fabric increases performance. Nevertheless, global exchanges on the wafer are still relatively fast, with edge-to-edge latency of around a microsecond.

Each core acts independently and executes its own code. Cores support nine concurrent threads. All datapath instructions may operate on registers and dataflow streams. Streams provide data  from memory, fabric, and inter-thread memory-backed FIFOs. Stream-descriptor registers offload loop-related overheads and set the stream length, access patterns, and termination conditions associated with the data flows. Hardware handles thread scheduling on a cycle-by-cycle basis. This enables task-driven asynchronous execution in the form of fine-grained processing initiated by data arrival. This asynchronous data flow inherently pipelines execution across tiles to mask latency and overlap communication with computation.

The design of the WSE removes the need and the associated overhead of an operating system and a message-passing library like MPI~\cite{mpi41}. The architectural elements open opportunities for novel algorithms and methods to speed up scientific calculations. While this requires a degree of reimagining application-to-hardware mappings, it can deliver significant performance payoffs.  Indeed, previous studies~\cite{WSE_HPC_1,WSE_HPC_2,WSE_HPC_3} demonstrate that the WSE is capable of delivering orders of magnitude higher performance than conventional systems for scientific workloads.

\subsection{Test Conditions}
We implemented the algorithm in the Tungsten programming language.
The language features a concise mathematical representation for core logic and task paralellism. A minimal version of the code (excluding debugging harness and atom re-assignment) is only 200 lines of code. Although the compiled high level code runs with high performance, we continued optimization with manual revisions to the assembly code produced as compiler output.

\begin{figure*}[t]
\centering
\newlength{\iw}
\settowidth{\iw}{\includegraphics{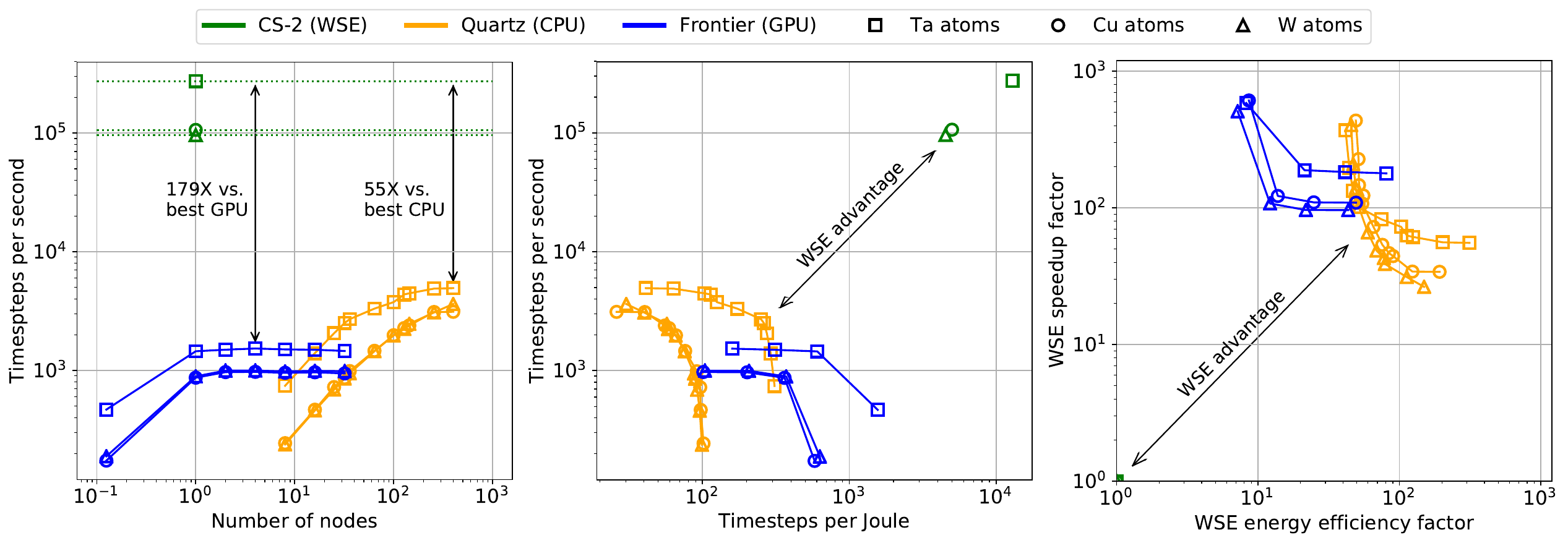}}
\subfloat[][]{\includegraphics[trim= 0 0 0.66\iw{} 0, clip=true, width=0.34\textwidth]{figures/performance-panel.pdf}\label{fig:perfpanela}}
\subfloat[][]{\includegraphics[trim= 0.34\iw{} 0 0.34\iw{} 0, clip=true, width=0.32\textwidth]{figures/performance-panel.pdf}\label{fig:perfpanelb}}
\subfloat[][]{\includegraphics[trim= 0.66\iw{} 0 0 0, clip=true, width=0.34\textwidth]{figures/performance-panel.pdf}\label{fig:perfpanelc}}
\caption{Measured performance and energy efficiency of single WSE compared to multi-node GPU and CPU systems for Ta, Cu, and W EAM benchmark simulations with 801,792 atoms. \textbf{(a)} For Ta, WSE (green square) achieved 179x and 55x speedup compared to the maximum simulation rates on GPU (blue squares) and CPU (orange squares) systems, respectively; \textbf{(b)} WSE also demonstrated one to two orders of magnitude improvement in energy efficiency over both CPU and GPU systems; \textbf{(c)} Relative energy efficiency and performance of CPU and GPU systems compared to WSE, showing Pareto front dominance of WSE on both metrics.} 
\label{fig:perfpanel}
\vspace*{-7pt}
\end{figure*}

We ran three types of simulations to assess performance and breadth of applicability for the algorithm:

\begin{enumerate}[wide]
\item Simulation domains appropriate for grain boundary problems (Table~\ref{tab:realistic}, Fig.~\ref{fig:perfpanel}). They use thin slab geometries $\sim(60\mathrm{nm} \times 60\mathrm{nm} \times 2\mathrm{nm})$ with open boundaries. Simulations use a single atom type (copper, tungsten, and tantalum) at room temperature for uniform crystal lattices. Each configuration was equilibrated in LAMMPS for 20k timesteps with a 2~fs timestep at 290~K before benchmarking. Our benchmark simulations have 800k atoms, using 94\% of the CS-2's cores. With over 10 atomic layers in thickness and hundreds in width, this size is relevant for grain boundary simulations, allowing observation of nucleation and evolution of multi-phase grain boundaries, and is big enough to directly compare to electron microscopy imaging~\cite{frolov2020}.

\item Performance measurements under highly controlled conditions (Table~\ref{tab:regression}).
Here, we run a parameter sweep of configurations, varying the independent variables, \ncandidate\ and \ninteraction. For these runs, the atoms' initial conditions are a regular 2D grid configuration. A neighborhood-size parameter determines the candidate count. The interaction cutoff threshold controls the number of interactions. Atoms hold their position throughout performance measurement because the time-stepping constant is zero. At the end of every timestep, the cores record a hardware clock cycle counter in a scratch memory buffer. We also conduct measurements over million-timestep intervals to verify that performance remains stable.

\item Simulations of grain boundaries, where distinct lattice organizations meet. Because atoms migrate due to structural dynamics of the grain boundary, we verify that online atom reassignment maintains low assignment costs.

\end{enumerate}

We collected reference performance data for GPU and CPU multi-node systems using the EAM implementation in the production LAMMPS code~\cite{LAMMPS}. Short LAMMPS simulations were run for three different realistic metallic crystals (Table~\ref{tab:realistic}). GPU performance was measured on OLCF's Frontier supercomputer. Frontier is currently the fastest supercomputer in the world and the first Exascale supercomputer, achieving 1.1 FP64 exaFLOPS with the LINPACK benchmark~\cite{top500}. Each Frontier node consists of 8 AMD Instinct MI250X GPU compute dies (GCDs) attached to an AMD Optimized 3rd Generation EPYC 64C 2GHz CPU\@. A total of 9,408 nodes are connected together by HPE's Slingshot-11 network.
CPU performance was measured on LLNL's Quartz cluster, which has 2.1 GHz Intel Xeon E5-2695 v4 (``Broadwell'') CPUs connected by an Omni-Path network~\cite{top500_Quartz}.

\begin{table*}[th]
\caption{Results for 800,000 atom models:
Predicted and measured performance (timesteps per second) on the WSE compared with Frontier(GPU) and Quartz(CPU)}
\label{tab:realistic}
\centering
\begin{tabular}{@{}llccccc|cccc@{}}
  \toprule
  \multicolumn{1}{@{}l}{\multirow{2}{*}{Element}}
  & \multicolumn{1}{l}{\multirow{2}{*}{Replication}}
  & \multicolumn{1}{l}{\multirow{2}{*}{Atoms}}
  & \multicolumn{1}{l}{Interactions/}
  %& \multicolumn{1}{l}{\multirow{2}{*}{Ca}}
  & \multicolumn{1}{l}{Predicted}
  & \multicolumn{1}{l}{Measured}
  & \multicolumn{1}{l}{Prediction}
  & \multicolumn{1}{|l}{Frontier}
  & \multicolumn{1}{l}{Quartz}
  & \multicolumn{1}{l}{WSE vs.}
 & \multicolumn{1}{l@{}}{WSE vs.}
  \\
  % Element
  & % Replication
  & % Atoms
  & \multicolumn{1}{l}{\enspace{Candidates}}
  & \multicolumn{1}{c}{(WSE)}
  & \multicolumn{1}{c}{(WSE)}
  & \multicolumn{1}{c}{(error)}
  & \multicolumn{1}{|c}{(GPU)}
  & \multicolumn{1}{c}{(CPU)}
  & \multicolumn{1}{l}{Frontier}
  & \multicolumn{1}{l@{}}{Quartz}
\\
  \midrule
  Copper (Cu)~\cite{copper}     & $174 \times 192 \times 6$ & 801,792 & 42/  224  & 104,895 & 106,313 & 1.3\% &   973 & 3,120 & \textbf{109x} & \textbf{34x} \\
  Tungsten (W)~\cite{tungsten}  & $256 \times 261 \times 6$ & 801,792 & 59/  224  &  93,048 &  96,140 & 3.2\% &   998 & 3,633 & \textbf{96x} & \textbf{26x}\\
  Tantalum (Ta)~\cite{tantalum} & $256 \times 261 \times 6$ & 801,792 & 14/   80  & 270,097 & 274,016 & 1.4\% & 1,530 & 4,938 & \textbf{179x} & \textbf{55x}\\
  \bottomrule
\end{tabular}
\vspace*{-5pt}
\end{table*}

Performance is measured in timesteps per second (higher is better).  WSE simulations measure performance running 8000 timesteps and counting the number of clock cycles. LAMMPS uses its internal timers output as ``Loop time'' in the LAMMPS log file. In all cases, this includes only time spent in the MD Verlet loop, and does not include initialization, setup, or finalization time, in order to calculate an accurate time-stepping rate independent of the number of timesteps run. However, initialization and finalization take less than 2 minutes for all the simulations run, which would give negligible ($<$ 0.2\%) overhead for a typical 24-hour production run.

The KOKKOS package in LAMMPS, which implements performance portability abstractions from the Kokkos library~\cite{kokkos} was used to run on the AMD GPUs using the HIP backend. GPU-aware MPI was used for multi-GPU runs on Frontier. LAMMPS simulations used FP64 (double) precision while the WSE implementation used FP32 (single) precision.

                % VI
\vspace{-8pt}
\section{Performance Results}
\label{section:results}

\subsection{Time to Solution and Strong Scaling}

To demonstrate strong scaling, we ran identical simulations of three metallic crystals on three different platforms. Fig.~\ref{fig:perfpanela} shows the achieved timesteps per second (and therefore the achievable simulated timescale per day).   Several important facts emerge immediately:
\begin{enumerate}[wide]
    \item GPUs scale poorly for systems of this size.   For one Frontier node having eight GCDs, the performance limit has been achieved; evidently on the order of 100,000 atoms per GPU is the limit to strong scaling. This is likely due to overheads for kernel launch, and other factors that force coarse parallelism granularity.
 
    \item   CPUs (Quartz system data) are amenable to scaling and finer parallel grain.  However, the scaling stalls at 400 dual-socket nodes, so 1000 atoms per CPU socket seems to be the limit. MPI communication costs are likely the limiter for this case.

    \item The picture is far more optimistic with wafer-scale systems which can strong scale to single atom per core grain, have small overheads, and have the communication capacity to tolerate this small parallel granularity.
\end{enumerate}

The tantalum simulation with 801,792 atoms on WSE runs 179x faster than LAMMPS on Frontier and 55x faster than LAMMPS on Quartz 
(Fig.~\ref{fig:perfpanela}, Table~\ref{tab:realistic}). This unprecedented speedup compared to the simulations on traditional platforms establishes the advantage of this work. Simulations of copper (109x speedup over GPU, 34x over CPU)  and tungsten  (96x, 26x) systems also validate the advantage with a 26x to 109x speedup over Frontier and Quartz, demonstrating generality.
The WSE also achieves 
one to two orders of magnitude better energy efficiency  
than Quartz and Frontier (Fig.~\ref{fig:perfpanelb}). At the limit of their strong scaling regimes, LAMMPS on Frontier and Quartz is highly \textit{energy inefficient}, as reflected by achieving fewer timesteps per Joule as it achieves more timesteps per second. As we add more nodes to the simulation, both timesteps per second and timesteps per Joule decrease (some data not plotted). Because the different metal systems impose different amounts of work, we normalized all WSE results to 1 and plot the WSE's energy and power advantage versus Frontier and Quartz (Fig.~\ref{fig:perfpanelc}).

It is notable here that CPUs (Quartz) are more effective than GPUs (Frontier), and that the GPU granularity cannot be less than about 400,000 atoms per node without disastrous loss of efficiency. We attribute this to significant overheads, including the large kernel launch latency on the GPU.  In comparison with Frontier node having 8 GCDs, the WSE achieves roughly 30-fold more timesteps per Joule, and that advantage grows as more GPU nodes are used, at ever larger power but with little improvement in performance.  Indeed, the data show the best GPU energy efficiency when using only one of the eight GCDs on a single Frontier node.

\subsection{Performance Analysis}

Theoretically, the algorithm's wall-clock run time \twall\ for a simulated timestep depends on the number of candidates (\ncandidate) and actual (\ninteraction) interactions per atom. To a much lesser extent, it also depends on the square root of the candidate count due to the construction of the neighborhood multicast. We test this empirically by fitting a linear model to a controlled sweep of performance measurements:
\begin{equation}\label{eq:linearmodel}
\twall=A\times\ncandidate+B\times\ninteraction+C
\end{equation}

Table~\ref{tab:regression} accurately accounts for observed timestep time, with $r^2=0.9998$. Cycle counts on the platform are remarkably stable. On a per-tile basis, the standard deviation of timestep time is 0.11\% ($3,477 \pm 3.77$~cycles). However, the tiles are locally synchronized with each neighborhood exchange region. When we first average the timestep times across the entire array, and then measure standard deviation it becomes 91 ppm ($3,477 \pm 0.316$~cycles).

The simple linear model also correctly matches performance of the true crystal simulations. In this case, the true crystal runs achieve about 1--3\% higher performance than expected. This may be due to atom motion transiently causing few than expected interactions for some atoms on some timesteps.

\begin{table}[tp]
\caption{Linear regression of time per timestep}
\label{tab:regression}
\centering
\begin{tabular}{@{}rrr@{}}
  \toprule
  Per candidate ($A$)
  & Per interaction ($B$)
  & Fixed ($C$)
  \\
  \midrule
  %Baseline  & 72.9 & 427.5 & 1367.9 \\
   26.6 ns & 71.4 ns & 574.0 ns \\
  \bottomrule
\end{tabular}
\vspace*{-5pt}
\end{table}

\subsection{Weak Scaling}
\begin{figure}[tp]
\centerline{\includegraphics[width=0.33\textwidth]{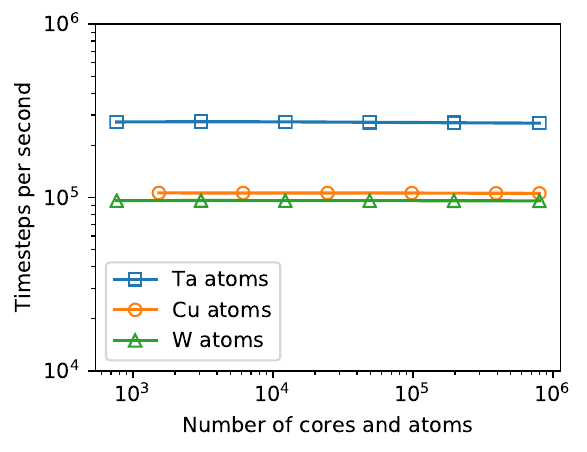}}
\vspace*{-7pt}
\caption{Perfect weak scaling results (within 1\%) across 3 orders of magnitude of core counts on a single WSE system.}
\label{fig:WSEweakscaling}
\vspace*{-7pt}
\end{figure}

Figure~\ref{fig:WSEweakscaling} reports weak-scaling results across \textit{three orders-of-magnitude core counts} on the WSE. The weak-scaling tests simultaneously increase the problem size and the number of WSE cores, always distributing one atom per core. This delivers perfect weak scaling on the wafer (to within 1\%) as core recruitment scales by three orders of magnitude.
Though not done in this work, distributing multiple atoms per core could further increase the problem size when all cores of the wafer are engaged \cite{NETLLDRD,NETLGitLab}.

\subsection{Peak Performance}
We analyze utilization as the ratio of algorithm-specified FLOPS to theoretical peak platform FLOPS.
Because MD includes diverse operations including spline lookups and elementary functions, Table~\ref{tab:flopmodel} shows our full accounting. We perform FLOP accounting in the basis $(\ninteraction, \ncandidate, 1)$ to obtain utilization for the three metallic crystals we simulated (Table~\ref{tab:utilization}).
We credit Frontier and Quartz runs with the same FLOP model even though this is slightly generous because LAMMPS foregoes most candidate processing by reusing neighbor lists across timesteps. Application initialization and finalization occurs in less than two minutes for all systems; for a typical 24 hour run this would deduct a negligible (less than quarter percent) utilization factor.

\begin{table}[tp]
\caption{FLOP count for all adds, muls, and other (e.g., conversion) steps. We convert to theoretical at-peak run time (based on processor clock) and compare with measured time to determine utilization of each algorithm component.}
\label{tab:flopmodel}
\begin{tabular}{@{}l r r r l @{}}
Term & $+$ & $\times$ & $\sim$ &  Note \\
\hline
%Candidate                                                     \\
$\vec{r_{ij}} \gets \vec{r_j} - \vec{r_i}$         & 3 &   & & Relative displacement      \\
$r_{ij}^2 \gets \vec{r_{ij}} \cdot \vec{r_{ij}}$   & 2 & 3 & & Squared distance   \\
$r_{ij}^2 < \threshold^2$                            & 1 &   & & Threshold check     \\
\textbf{Per Candidate Subtotal}                                 & 6 & 3 &  & \textbf{5.3 ns / 26.6 ns = 20\%}  \\
\hline
%Interaction   \\
${r_{ij}}^{-1} \gets ({r_{ij}}^2)^{-1/2}$                                            & 3 & 8 & 1 & Newton-Raphson \\
$r_{ij} \gets {r_{ij}}^2 {r_{ij}}^{-1}$                                              &   & 1 &   & Euclidean distance \\
%$x \gets \alpha r_{ij}$                                                             &   & 1       \\
%$\bar{x} \gets \left \lfloor x \right \rfloor $                                     &   &   & 1   \\
%$\Delta{x} \gets x - \bar{x}$                                                       & 1           \\
%$k \gets int(x)$                                                                    &   &   & 1   \\
$k,\Delta{x} \gets \mathrm{segment}(r_{ij})$                                                  & 1 & 1 & 2 & Spline segment  \\
$\sum_j \rho[k](\Delta{x})$                                                          & 3 & 2 &   & Density evaluation \\
% $\rho_j[k](\Delta{x})$                                                             & 0 & 0 &   & Optimized out    \\
$\rho'[k](\Delta{x}), \phi'[k](\Delta{x})$                                           & 2 & 2 &   & Linear splines   \\
% $\rho'_j[k](\Delta{x})$                                                            & 0 & 0 &   & Optimized out    \\
%$$                                                                & 1 & 1 &   & Linear spline    \\
$\sum_j ((F'_i + F'_j) \rho'(\cdot) + \phi'(\cdot)) r_{ij}^{-1} \vec{r_{ij}}$        & 5 & 5 &   & Force evaluation \\
\textbf{Per Interaction Subtotal}                                                        &14 &19 & 3 & \textbf{21.2ns / 71.4ns = 30\%} \\
\hline
%Particle \\
$k,\Delta(x) \gets \mathrm{segment}(\rho_i)$                                                  & 1 & 1 & 2 & Spline segment  \\
$F'_i[k](\Delta{x})$                                                                 & 1 & 1 &   & Embedding component   \\
$\int{\vec{v_i}}, \int{\vec{r_i}}$                                                   & 6 &   &   & Verlet integration \\
\textbf{Fixed Subtotal }                                                          & 8 & 2 & 2 & \textbf{7.1ns / 574ns = 1\%} \\
\hline
\end{tabular}
\vspace*{-5pt}
\end{table}

\begin{table}[tp]
    \caption{Utilization (Fraction of Peak) for three different architectures.}
    \label{tab:utilization}
    \centering
    \begin{tabular}{l|r S|S S S}
         Machine     & Chips                & \multicolumn{1}{c|}{Peak}     &  Cu     & W      & Ta   \\
                   &                      & \multicolumn{1}{c|}{PFLOP/s}  &  \multicolumn{3}{c}{ Utilization (\%) }\\
             \hline
          CS-2     &     1 WSE               & 1.45          & \textbf{22.}  & \textbf{23.}  & \textbf{20.}  \\
        Frontier    &    32 GCD               & 0.77          &  0.4 &  0.4 &  0.2 \\
        Quartz & 800 CPU & 0.50           &  1.9 &  2.5 &  1.0 \\
    \end{tabular}
\vspace*{-5pt}
\end{table}

\subsection{Grain Boundary Problem}
\begin{figure}[t]
\centerline{\includegraphics[width=0.5\textwidth]{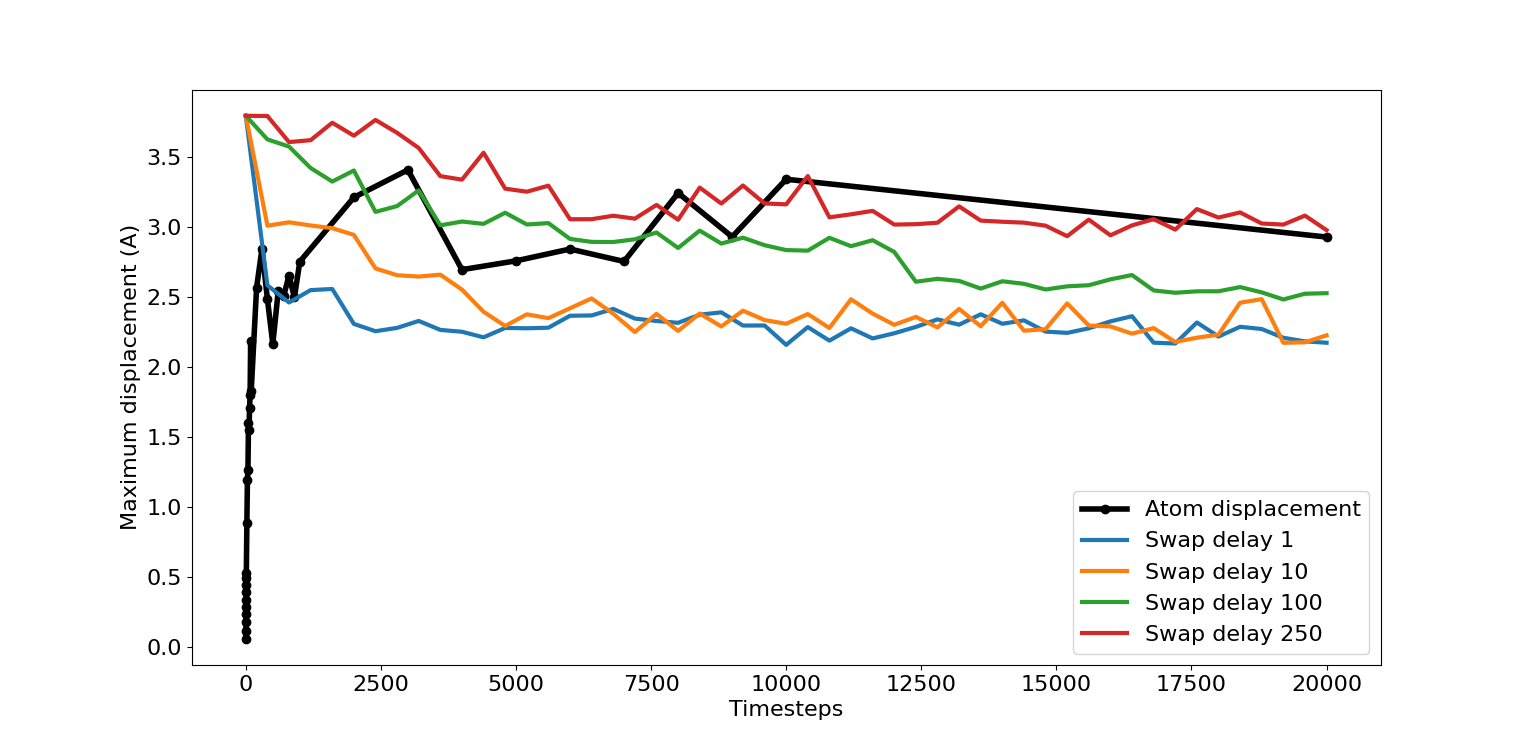}\vspace*{-1em}}
\caption{Atom motion and assignment cost. (black) Largest max-norm of any atom displacement in the X-Y plane, as a function of time. Linear motion of atoms (max norm) over time. (colored lines) Assignment cost over time, as a function of how often swaps are attempted. The legend lists the delay in timesteps between swap attempts.}
\label{fig:sorting}
\vspace*{-7pt}
\end{figure}
The grain boundary problem has the same characteristics as the thin slab geometries except it requires an occasional additional step to reassign the atom-to-core mapping as the atoms diffuse over time.
We perform atom swaps (Sec.~\ref{sec:atomswap}) at regular intervals to maintain an efficient atoms-to-core mapping. For the grain boundary problem (Fig.~\ref{fig:GB}) we explored the effect of the atom swap interval. The black line in Fig.~\ref{fig:sorting} shows the max norm of the $x$ and $y$~displacements over time. 
This time evolution will be roughly similar regardless of the time starting point. The colored lines show the atom-to-core assignment cost evolution, starting from a sub-optimal initial mapping. The different colors represent  swap intervals from 1 to 250 timesteps. After an initial transient the atom swaps maintain the neighborhood exchange distance to within 3\AA~plus the EAM cutoff distance, for swap intervals of 100 timesteps or less. As a comparison, our best off-line attempt at optimizing of the assignment cost yielded 2.1\AA~plus the EAM cutoff. We used 62,500 cores for 61,600 atoms, and left 900 empty. A swap takes roughly the same time as a timestep.  Fig.~\ref{fig:sorting} suggests that swapping atoms every 10--100 steps is sufficient for maintaining low assignment cost with only modest additional overhead.

\subsection{Fabric Load with Periodic Boundaries}
\label{sec:FabricLoadPBC}
To demonstrate completeness, we also evaluate periodic boundary conditions.
Periodic boundary conditions double the fabric data transfer amounts during neighborhood multicast, because logical core neighbors are two hops away (Fig.~\ref{fig:periodic}, Sec.~\ref{sec:pbc}).

The fabric routers can route data simultaneously in both directions while also delivering data to the core. This means the lateral transfer bandwidth can cover the extra load imposed by PBCs without slowdown. We measured the performance of the position exchange with and without PBCs, and verified that they indeed take the same amount of time. However periodicity is not free; there is still a computational cost to do modular arithmetic in the distance calculation. A production implementation of the algorithm could avoid the modular arithmetic by intercepting traffic crossing a boundary and offsetting the coordinate by the domain width.
                    % VII
  \vspace{-5pt}
\subsection{Methodology for High Performance}
\label{section:optimization}
Achieving high performance code began with developing an architecture-specific performance model that accounts cycles in each phase of the computation. We used this model to assess the implications of various design choices during the mapping stage. After we arrived at a general architectural direction, this model became more detailed to explore the implications of specific code paths and execution strategies.

\begin{figure}[tp]
\includegraphics[width=0.5\textwidth]{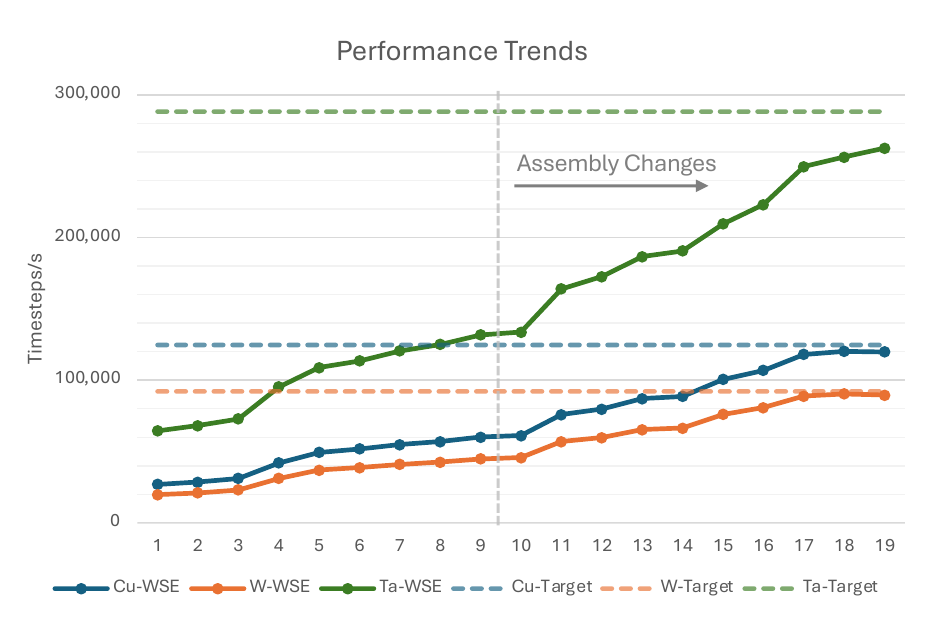}
\vspace*{-25pt}
\caption{Performance of material simulations across code changes, showing extensive optimization effort.}
\label{fig:performancetrends}
\vspace*{-7pt}
\end{figure}

Our first functioning EAM code was 5.6x slower than the performance model prediction.
We used the linear regression methodology (Sec.~\ref{section:results}) to assess every algorithmic step. This provided a map of where to find and fix performance gaps. Fig.~\ref{fig:performancetrends} shows measured performance after each code change.

The first group of changes were made in the high-level, domain-specific language called Tungsten. Optimizations at this level include loop vectorization, eliminating support for unused features, interleaving some terms in memory layout, and minimizing use of conditional logic. This brought the implementation within 2x of the goal to match the performance model.
The remaining optimizations were done by manually editing the compiler's assembly output. This includes instruction reordering to avoid pipeline stalls, reusing stream descriptors, shifting array offsets to avoid bank conflicts, and leveraging hardware offloads in cases the compiler missed.          % VII-F
\section{Implications}
\label{section:implications}

\begin{table}[tp]
\caption{Projected Performance Gains from Future Optimizations}
\centering
\begin{tabular}{@{}
    l
    S[table-format=1.0]
    S[table-format=2.1]
    S[table-format=2.0]
    S[table-format=3.0]
    r
    r
    r
    @{}
}
  \toprule
  \multicolumn{1}{@{}l}{\multirow{2}{*}{Description}}
  & \multicolumn{1}{|c}{Mcast}
  & \multicolumn{1}{c}{Miss}
  & \multicolumn{1}{c}{Interaction}
  & \multicolumn{1}{c}{Fixed}
    & \multicolumn{1}{|r}{Ta}
  & \multicolumn{1}{r}{W}
  & \multicolumn{1}{r@{}}{Cu} \\
  %& \multicolumn{1}{r}{W}
  %& \multicolumn{1}{r@{}}{Cu}
  & \multicolumn{4}{|c}{nanoseconds (\% scale)}
  %& \multicolumn{1}{c}{(ns)}
  %& \multicolumn{1}{c}{(ns)}
  %& \multicolumn{1}{c}{(ns)}
  & \multicolumn{3}{|c}{1,000 timestep/s } \\
  \midrule
  Baseline             &   6               & 21               & 92              & 574              &  270   & 130 & 150 \\
  Fixed cost       &   6               & 21               & 92              & 287\unit{(50\%)} &  290   & 130 & 160 \\
  Neighbor list        &   6               & 2.1\unit{(10\%)} & 92              & 287              &  460   & 140 & 180 \\
  Symmetry             &   6               & 2.1              & 46\unit{(50\%)} & 287              &  650   & 230 & 270 \\
  Parallel             &   3\unit{(50\%)}  & 1.0\unit{(50\%)} & 23\unit{(50\%)} & 287              &  1,100 & 430 & 510 \\
  \bottomrule
\end{tabular}
\label{table:projectedperf}
\vspace*{-5pt}
\end{table}

\begin{table*}[tb]
\caption{Modeled multi-wafer performance as a function of the ghost region size. Between 92\% and 99\% of the single-wafer performance is preserved. }
\label{tab:multiwafer}
\centering
\begin{tabular}{@{}lccccc|cccc|cccc@{}}
  \toprule
\multicolumn{1}{@{}l}{\multirow{2}{*}{Element}} & \multicolumn{1}{l}{\multirow{2}{*}{$X$}} & \multicolumn{1}{l}{\multirow{2}{*}{$Z$}} & \multicolumn{1}{l}{\multirow{2}{*}{$\Nnode$}} & \multicolumn{1}{l}{\multirow{2}{*}{$\frac{\threshold}{r_\mathrm{lattice}}$}} & \multicolumn{1}{l}{\multirow{2}{*}{$t_\mathrm{wall}$ ($\mu$s)}} & \multicolumn{4}{|c}{Low Utilization ($\Ninterior/\Nnode = \text{20\%}$)} & \multicolumn{4}{|c@{}}{High Utilization ($\Ninterior/\Nnode = \text{80\%}$)} \\
\multicolumn{6}{c}{} & \multicolumn{1}{|c}{$\lambda$} & \multicolumn{1}{c}{$k$} & \multicolumn{1}{c}{Timesteps/s} & \multicolumn{1}{c}{Performance} & \multicolumn{1}{|c}{$\lambda$} & \multicolumn{1}{c}{$k$} & \multicolumn{1}{c@{}}{Timesteps/s} & \multicolumn{1}{c}{Performance}\\
  \midrule
Cu&	283&	10&	800,890&	1.94&	9.41&	78&	20&	105,152 & 99\% &	15&	3&	99,239 & 93\% \\
W&	317&	8&	803,912&	2.02&	10.4&	88&	21&	95,281 & 99\% &	17&	4&	91,743 & 95\% \\
Ta&	317&	8&	803,912&	1.39&	3.65&	88&	31&	269,214 & 98\% &	17&	6&	251,046 & 92\% \\
  \bottomrule
\end{tabular}
\vspace*{-5pt}
\end{table*}

\subsection{Additional Acceleration}
\label{section:futureoptimization}

The paradigm we developed offers possibilities to accelerate MD beyond 1M timestep/s. Starting with the baseline performance model (Table~\ref{tab:regression}), we attribute 6 ns per candidate for neighborhood multicast and re-express the model in a new basis (Table~\ref{table:projectedperf}). We conservatively project performance boost of four independent optimizations:

\begin{enumerate}[wide]
\item \textbf{Reduce fixed cost.} In Section \ref{section:optimization} we described targeted optimizations to reduce the per candidate and per timestep costs. These resulted in 5x cost improvements. Minimal effort was focused on reducing the fixed costs. It is safe to assume another 2x speedup can be achieved for the fixed cost component.

\item \textbf{Neighbor List.} Other MD implementations reuse their neighbor lists (Section \ref{section:performance_criteria}) for tens of timesteps. In current runs almost half of runtime is spent processing rejected candidates. We conservatively model that the neighbor list can be re-examined every tenth timestep.

\item \textbf{Force Symmetry.}
We can calculate $(\cdot)_{ij}$-terms when $i<j$ and send results from $i$ to $j$. Cropping neighborhood multicast to the right sub-neighborhood accomplishes the restriction.
Bandwidth considerations make it impractical to unicast back to the originating worker. Instead, a neighborhood reduction 
operates as the reverse of neighborhood multicast. The reverse step of multicast forwarding is naturally a 2:1 sum reduction performed directly at the branch. The reduction retains the multicast's systolic dataflow properties.

\item \textbf{Multi-core workers.} Several strategies are available for spreading a worker's workload over multiple cores. One strategy is to use separate cores to compute the $\phi$ and $\rho$ force components. Another strategy is to load balance interactions among cores, e.g.,~by alternating assignment of even versus odd incoming atoms among a pair of cores. We conservatively assume that with 4-core parallelization a 2x cost reduction is possible on multicast, reject, and interaction processing.
\end{enumerate}

As shown in Table \ref{table:projectedperf}, with all of these optimizations combined we conservatively project performance in excess of one million timesteps for the tantalum benchmark.

\subsection{Molecular Dynamics}

Traditional exascale architectures do not offer a clear path to $\sim$100 microseconds and millisecond atomistic MD simulations of materials and chemical systems.  Achieving these extended timescales will be transformational for many areas of scientific research including designing materials for nuclear energy, thermally activated catalytic reactions, phase nucleation close to equilibrium, and protein folding.
In particular, the future of high-intensity decarbonized energy generation technologies like nuclear fusion require materials science advances that will be driven by MD simulations at the unprecedented $\sim$100 microsecond timescale
achieved here.

Our work demonstrates that novel wafer-scale computer architectures can achieve a major increase in the maximum simulation rate of complex atomistic systems. This opens opportunities to greatly accelerate the rate of 
scientific discovery in many different areas,
 particularly the design of new materials with enhanced performance characteristics that are
 needed to address global societal challenges such as climate change. The mapping of the EAM interatomic potential to the WSE can be readily adopted for other MD potentials,
including both empirical potentials ~\cite{MEAM, LJ, tersoff}, as well as high accuracy machine-learning potentials \cite{deepmd_GB,carbon_GB,allegro_GB}. 
We anticipate the next generations of wafer-scale technology will continue this trend of breaking through the long-standing timescale barrier.

\subsection{Weak Scaling to Many Wafers}

To weak scale the dataflow EAM MD algorithm to many WSE nodes, we can distribute non-overlapping subdomains to individual nodes. Each node holds unique \emph{interior} atoms and aliased \emph{ghost} atoms. The ghosts exist in an expansion of a node's subdomain by $\lambda$ lattice units beyond its boundary.

A simple multi-node execution model shows the feasibility of weak scaling. Let each node have $\Nnode=\Ninterior+\Nghost$ atoms. At the start of a \emph{time period} all $\Nnode$ are valid. Each simulation timestep invalidates the outermost $2\threshold$-wide strip of ghosts. Therefore, the node may run $k=\lambda{r_\mathrm{lattice}}/2\threshold$ timesteps. After this, it must receive 192-bits per ghost of refreshed position and velocity information.

Transmission time to update ghost atoms depends on the bandwidth $\omega$ and inter-node latency $\tau$. Current generation WSE achieve $\omega=1.2 \mathrm{Tb/s}$; exascale interconnects achieve $\tau=2 \mathrm{\mu{s}}$.

For a thin slab subdomain with $\Ninterior=({X^2}\times{Z})$, $\Nnode=(X+2\lambda)^2\,Z$, the time to compute a period of $k$ timesteps is $t_\mathrm{period}=k\,t_\mathrm{wall}+\tau+192\,\Nghost/\omega$.

Useful scaling typically occurs when $\Ninterior$ is between 20\% and 80\% of the total number of atoms in a node. Greater ghost counts achieve higher timestep/s by amortizing away transmission latency; this comes at the cost of smaller subdomains.

Table~\ref{tab:multiwafer} shows the modeled performance of the multi-wafer mapping. More than 92\% of the single-wafer performance in Table~\ref{tab:realistic} is preserved. Increasing the size of the ghost region further improves the performance at the cost of hurting utilization.
With this model, we estimate the already deployed 64-node WSE clusters could simulate tantalum systems of over 10 or 40 million atoms at 269k or 251k timesteps/s, respectively.

\subsection{High-performance Computing}

GPU-accelerated supercomputer architectures excel at problems requiring immense computational throughput such as dense linear algebra and weak-scaling MD\@.  Alas, such throughput-based architectures are a mismatch to the challenge of breaking the MD timescale barrier, which is a strong-scaling problem and therefore extremely latency-sensitive.  Our approach integrates algorithms, software, and hardware to address the problem of poor processor utilization induced by conventional architectures' relatively high interprocessor communication costs.  We demonstrate revolutionary benefits both in terms of time to solution and energy efficiency and do so fully independently of any custom or domain-specific hardware.  Our work prescribes a radical path forward from exascale to the next era of extreme-scale computing, in which applications whose performance otherwise would stagnate can be accelerated.  Future work will focus on extending the strong-scaling efficiency demonstrated here to facility-level deployments, potentially leading to an even greater paradigm shift in the Top500 supercomputer list than that introduced by the GPU revolution.
               % VIII

\section*{Acknowledgments}
\small{Sandia National Laboratories is a multimission laboratory managed and operated by National Technology and Engineering Solutions of Sandia, LLC., a wholly owned subsidiary of Honeywell International, Inc., for the U.S. Department of Energy’s National Nuclear Security Administration under contract DE-NA-0003525. The SNL document release number is SAND2024-05460O. Los Alamos National Laboratory is operated by Triad National Security, LLC, for the National Nuclear Security Administration of U.S. Department of Energy (Contract No. 89233218CNA000001). The LANL document release number is LA-UR-24-23572. This work was performed under the auspices of the U.S. Department of Energy by Lawrence Livermore National Laboratory under Contract DE-AC52-07NA27344. The LLNL document release number is LLNL-CONF-863696.
The authors thank Ryan Humble for his contributions to the initial implementation specification.}

\bibliographystyle{IEEEtran}
\bibliography{references}

\begin{thebibliography}{10}
\providecommand{\url}[1]{#1}
\csname url@samestyle\endcsname
\providecommand{\newblock}{\relax}
\providecommand{\bibinfo}[2]{#2}
\providecommand{\BIBentrySTDinterwordspacing}{\spaceskip=0pt\relax}
\providecommand{\BIBentryALTinterwordstretchfactor}{4}
\providecommand{\BIBentryALTinterwordspacing}{\spaceskip=\fontdimen2\font plus
\BIBentryALTinterwordstretchfactor\fontdimen3\font minus
  \fontdimen4\font\relax}
\providecommand{\BIBforeignlanguage}[2]{{%
\expandafter\ifx\csname l@#1\endcsname\relax
\typeout{** WARNING: IEEEtran.bst: No hyphenation pattern has been}%
\typeout{** loaded for the language `#1'. Using the pattern for}%
\typeout{** the default language instead.}%
\else
\language=\csname l@#1\endcsname
\fi
#2}}
\providecommand{\BIBdecl}{\relax}
\BIBdecl

\bibitem{carbon_GB}
K.~Nguyen-Cong, J.~T. Willman, S.~G. Moore, A.~B. Belonoshko, R.~Gayatri,
  E.~Weinberg, M.~A. Wood, A.~P. Thompson, and I.~I. Oleynik, ``Billion atom
  molecular dynamics simulations of carbon at extreme conditions and
  experimental time and length scales,'' in \emph{Proceedings of the
  International Conference for High Performance Computing, Networking, Storage
  and Analysis}, ser. SC '21.\hskip 1em plus 0.5em minus 0.4em\relax New York,
  NY, USA: Association for Computing Machinery, 2021.

\bibitem{tchipev2019twetris}
N.~Tchipev, S.~Seckler, M.~Heinen, J.~Vrabec, F.~Gratl, M.~Horsch,
  M.~Bernreuther, C.~W. Glass, C.~Niethammer, N.~Hammer \emph{et~al.},
  ``Twetris: Twenty trillion-atom simulation,'' \emph{The International Journal
  of High Performance Computing Applications}, vol.~33, no.~5, pp. 838--854,
  2019.

\bibitem{deepmd_GB}
W.~Jia, H.~Wang, M.~Chen, D.~Lu, L.~Lin, R.~Car, W.~E, and L.~Zhang, ``Pushing
  the limit of molecular dynamics with ab initio accuracy to 100 million atoms
  with machine learning,'' in \emph{Proceedings of the International Conference
  for High Performance Computing, Networking, Storage and Analysis}, ser. SC
  '20.\hskip 1em plus 0.5em minus 0.4em\relax IEEE Press, 2020.

\bibitem{allegro_GB}
B.~Kozinsky, A.~Musaelian, A.~Johansson, and S.~Batzner, ``Scaling the leading
  accuracy of deep equivariant models to biomolecular simulations of realistic
  size,'' in \emph{Proceedings of the International Conference for High
  Performance Computing, Networking, Storage and Analysis}, ser. SC '23.\hskip
  1em plus 0.5em minus 0.4em\relax New York, NY, USA: Association for Computing
  Machinery, 2023.

\bibitem{Plimpton2012}
S.~J. Plimpton and A.~P. Thompson, ``{Computational aspects of many-body
  potentials},'' \emph{MRS Bull.}, vol.~37, pp. 513--521, 2012.

\bibitem{frolov2018}
T.~Frolov, W.~Setyawan, R.~Kurtz, J.~Marian, A.~R. Oganov, R.~E. Rudd, and
  Q.~Zhu, ``{Grain boundary phases in bcc metals},'' \emph{Nanoscale}, vol.~10,
  pp. 8253--8268, 2018.

\bibitem{GrainBoundary}
T.~Frolov, D.~L. Olmsted, M.~Asta, and Y.~Mishin, ``Structural phase
  transformations in metallic grain boundaries,'' \emph{Nature Communications},
  vol.~4, no.~1, p. 1899, 2013.

\bibitem{LJ}
J.~E. Jones, ``On the determination of molecular fields. {I}. {F}rom the
  variation of the viscosity of a gas with temperature,'' \emph{Proceedings of
  the Royal Society of London A: Mathematical, Physical and Engineering
  Sciences}, vol. 106, no. 738, pp. 441--462, 1924.

\bibitem{EAM1}
M.~S. Daw and M.~Baskes, ``Semiempirical, quantum mechanical calculation of
  hydrogen embrittlement in metals,'' \emph{Physical Review Letters}, vol.~50,
  no.~17, p. 1285 – 1288, 1983, cited by: 2480.

\bibitem{EAM2}
M.~S. Daw and M.~I. Baskes, ``Embedded-atom method: Derivation and application
  to impurities, surfaces, and other defects in metals,'' \emph{Phys. Rev. B},
  vol.~29, pp. 6443--6453, Jun 1984.

\bibitem{EAM3}
S.~Foiles, M.~Baskes, and M.~Daw, ``Embedded-atom-method functions for the fcc
  metals {C}u, {A}g, {A}u, {N}i, {P}d, {P}t, and their alloys,'' \emph{Physical
  Review B}, vol.~33, no.~12, p. 7983 – 7991, 1986, cited by: 4054.

\bibitem{EAM4}
M.~S. Daw, S.~M. Foiles, and M.~I. Baskes, ``The embedded-atom method: a review
  of theory and applications,'' \emph{Materials Science Reports}, vol.~9,
  no.~7, pp. 251--310, 1993.

\bibitem{LAMMPS}
A.~P. Thompson, H.~M. Aktulga, R.~Berger, D.~S. Bolintineanu, W.~M. Brown,
  P.~S. Crozier, P.~J. {in 't Veld}, A.~Kohlmeyer, S.~G. Moore, T.~D. Nguyen,
  R.~Shan, M.~J. Stevens, J.~Tranchida, C.~Trott, and S.~J. Plimpton,
  ``{LAMMPS}---a flexible simulation tool for particle-based materials modeling
  at the atomic, meso, and continuum scales,'' \emph{Computer Physics
  Communications}, vol. 271, pp. 108\,171:1--34, Feb. 2022.

\bibitem{kernel_fusion_PR}
LAMMPS, ``Github pull request,''
  \url{https://github.com/lammps/lammps/pull/3758}, 2023.

\bibitem{PRD}
D.~Perez, B.~P. Uberuaga, and A.~F. Voter, ``The parallel replica dynamics
  method---coming of age,'' \emph{Computational Materials Science}, vol. 100,
  pp. 90--103, 2015, special Issue on Advanced Simulation Methods.

\bibitem{morimoto2021hardware}
G.~Morimoto, Y.~M. Koyama, H.~Zhang, T.~S. Komatsu, Y.~Ohno, K.~Nishida,
  I.~Ohmura, H.~Koyama, and M.~Taiji, ``Hardware acceleration of
  tensor-structured multilevel {E}wald summation method on {MDGRAPE-4A}, a
  special-purpose computer system for molecular dynamics simulations,'' in
  \emph{Proceedings of the International Conference for High Performance
  Computing, Networking, Storage and Analysis}, 2021, pp. 1--15.

\bibitem{anton_GB}
D.~E. Shaw~et al., ``Anton 3: twenty microseconds of molecular dynamics
  simulation before lunch,'' in \emph{Proceedings of the International
  Conference for High Performance Computing, Networking, Storage and Analysis},
  ser. SC '21.\hskip 1em plus 0.5em minus 0.4em\relax New York, NY, USA:
  Association for Computing Machinery, 2021.

\bibitem{gromacs}
{GROMACS development team}, \emph{{GROMACS} Documentation, Release 2024.1},
  Feb.~28, 2024, available from
  \url{https://manual.gromacs.org/current/manual-2024.1.pdf}.

\bibitem{gromacs2}
M.~J. Abraham, T.~Murtola, R.~Schulz, S.~Páll, J.~C. Smith, B.~Hess, and
  E.~Lindahl, ``Gromacs: High performance molecular simulations through
  multi-level parallelism from laptops to supercomputers,'' \emph{SoftwareX},
  vol. 1-2, pp. 19--25, 2015.

\bibitem{WSE_HPC_2}
K.~Rocki, D.~V. Essendelft, I.~Sharapov, R.~Schreiber, M.~Morrison,
  V.~Kibardin, A.~Portnoy, J.~Dietiker, M.~Syamlal, and M.~James, ``Fast
  stencil-code computation on a wafer-scale processor,'' in \emph{Proceedings
  of the International Conference for High Performance Computing, Networking,
  Storage and Analysis, {SC} 2020, Virtual Event / Atlanta, Georgia, USA,
  November 9-19, 2020}, C.~Cuicchi, I.~Qualters, and W.~T. Kramer, Eds.\hskip
  1em plus 0.5em minus 0.4em\relax {IEEE/ACM}, 2020, p.~58.

\bibitem{HotChips}
S.~Lie, ``Cerebras architecture deep dive: First look inside the
  hardware/software co-design for deep learning,'' \emph{{IEEE} Micro},
  vol.~43, no.~3, pp. 18--30, 2023.

\bibitem{mpi41}
\BIBentryALTinterwordspacing
{Message Passing Interface Forum}, \emph{{MPI}: A Message-Passing Interface
  Standard Version 4.1}, Nov.~2, 2023. [Online]. Available:
  \url{https://www.mpi-forum.org/docs/mpi-4.1/mpi41-report.pdf}
\BIBentrySTDinterwordspacing

\bibitem{WSE_HPC_1}
M.~Woo, T.~Jordan, R.~Schreiber, I.~Sharapov, S.~Muhammad, A.~Koneru, M.~James,
  and D.~V. Essendelft, ``Disruptive changes in field equation modeling: {A}
  simple interface for wafer scale engines,'' \emph{CoRR}, vol. abs/2209.13768,
  2022.

\bibitem{WSE_HPC_3}
H.~Ltaief, Y.~Hong, L.~Wilson, M.~Jacquelin, M.~Ravasi, and D.~E. Keyes,
  ``Scaling the ``memory wall'' for multi-dimensional seismic processing with
  algebraic compression on {C}erebras {CS-2} systems,'' in \emph{Proceedings of
  the International Conference for High Performance Computing, Networking,
  Storage and Analysis, {SC} 2023, Denver, CO, USA, November 12-17, 2023},
  D.~Arnold, R.~M. Badia, and K.~M. Mohror, Eds.\hskip 1em plus 0.5em minus
  0.4em\relax {ACM}, 2023, pp. 6:1--6:12.

\bibitem{frolov2020}
T.~Meiners, T.~Frolov, R.~E. Rudd, G.~Dehm, and C.~H. Liebscher,
  ``{Observations of grain-boundary phase transformations in an elemental
  metal},'' \emph{Nature}, vol. 579, pp. 375--378, 2020.

\bibitem{top500}
Top500, \url{https://www.top500.org/lists/top500/list/2023/11/}, [Online;
  accessed 25-March-2024].

\bibitem{top500_Quartz}
------, \url{https://www.top500.org/site/48247/}, [Online; accessed
  2-April-2024].

\bibitem{copper}
\BIBentryALTinterwordspacing
J.~B. Adams, S.~M. Foiles, and W.~G. Wolfer, ``Self-diffusion and impurity
  diffusion of fcc metals using the five-frequency model and the embedded atom
  method,'' \emph{Journal of Materials Research}, vol.~4, no.~1, pp. 102--112,
  1989. [Online]. Available: \url{https://doi.org/10.1557/JMR.1989.0102}
\BIBentrySTDinterwordspacing

\bibitem{tungsten}
X.~Zhou, H.~Wadley, R.~Johnson, D.~Larson, N.~Tabat, A.~Cerezo,
  A.~Petford-Long, G.~Smith, P.~Clifton, R.~Martens, and T.~Kelly, ``Atomic
  scale structure of sputtered metal multilayers,'' \emph{Acta Materialia},
  vol.~49, no.~19, pp. 4005--4015, 2001.

\bibitem{tantalum}
Y.~Li, D.~J. Siegel, J.~B. Adams, and X.-Y. Liu, ``Embedded-atom-method
  tantalum potential developed by the force-matching method,'' \emph{Phys. Rev.
  B}, vol.~67, p. 125101, Mar 2003.

\bibitem{kokkos}
C.~R. Trott, D.~Lebrun-Grandié, D.~Arndt, J.~Ciesko, V.~Dang, N.~Ellingwood,
  R.~Gayatri, E.~Harvey, D.~S. Hollman, D.~Ibanez, N.~Liber, J.~Madsen,
  J.~Miles, D.~Poliakoff, A.~Powell, S.~Rajamanickam, M.~Simberg,
  D.~Sunderland, B.~Turcksin, and J.~Wilke, ``Kokkos 3: Programming model
  extensions for the exascale era,'' \emph{IEEE Transactions on Parallel and
  Distributed Systems}, vol.~33, no.~4, pp. 805--817, 2022.

\bibitem{NETLLDRD}
W.~Saidi, W.~Shi, and D.~Van~Essendelft, ``{Materials Modeling on the Wafer
  Scale Engine},'' {LDRD-Prime, NETL}, 2023.

\bibitem{NETLGitLab}
D.~Van~Essendelft, T.~Jordan, M.~Woo, W.~Shi, L.~Chong, A.~Zidane, and H.~Kim,
  ``{The Wafer Scale Engine, Field Equation, Application Programming
  Interface},'' \url{https://mfix.netl.doe.gov/gitlab/tjordan/cerebrasdev},
  {NETL}, 2023.

\bibitem{MEAM}
M.~I. Baskes, ``Modified embedded-atom potentials for cubic materials and
  impurities,'' \emph{Phys. Rev. B}, vol.~46, pp. 2727--2742, Aug 1992.

\bibitem{tersoff}
J.~Tersoff, ``New empirical approach for the structure and energy of covalent
  systems,'' \emph{Phys. Rev. B}, vol.~37, pp. 6991--7000, Apr 1988.

\end{thebibliography}

\end{document}